\documentclass[conference]{IEEEtran}
\IEEEoverridecommandlockouts
\usepackage{cite}
\usepackage{amsmath,amssymb,amsfonts,cuted}
\usepackage{algorithmic}
\usepackage{graphicx}
\usepackage{textcomp}
\usepackage{xcolor}

\usepackage{multirow}
\usepackage{makecell}
\usepackage{stfloats}
\usepackage{graphicx}
\usepackage{pgffor} % 提供循环功能
\usepackage{array}
\usepackage{booktabs}
\usepackage{hyperref}
\usepackage{orcidlink}

\def\BibTeX{{\rm B\kern-.05em{\sc i\kern-.025em b}\kern-.08em
    T\kern-.1667em\lower.7ex\hbox{E}\kern-.125emX}}
\begin{document}

\title{Joint Generalized Cosine Similarity: A Novel Method for N-Modal Semantic Alignment Based on Contrastive Learning}

\author{\IEEEauthorblockN{1\textsuperscript{st} Yiqiao Chen}
\IEEEauthorblockA{\textit{College of Global Talents} \\
\textit{Beijing Institute of Technology, Zhuhai}\\
Zhuhai, China \\
565196860@qq.com}
\and
\IEEEauthorblockN{1\textsuperscript{st} Zijian Huang}
\IEEEauthorblockA{\textit{Faculty of Frontier Sciences} \\
\textit{Harbin Institute of Technology, Shenzhen}\\
Shenzhen, China \\
~\orcidlink{0000-0001-6849-8827}}
}

\maketitle

\begin{abstract}
Alignment remains a crucial task in multi-modal deep learning, and contrastive learning has been widely applied in this field. However, when there are more than two modalities, existing methods typically calculate pairwise loss function and aggregate them into a composite loss function for the optimization of model parameters. This limitation mainly stems from the drawbacks of traditional similarity measurement method (i.e. they can only calculate the similarity between two vectors). To address this issue, we propose a novel similarity measurement method: the Joint Generalized Cosine Similarity (JGCS). Unlike traditional pairwise methods (e.g., dot product or cosine similarity), JGCS centers around the angle derived from the Gram determinant. To the best of our knowledge, this is the first similarity measurement method capable of handling tasks involving an arbitrary number of vectors. Based on this, we introduce the corresponding contrastive learning loss function , GHA Loss, and the new inter-modal contrastive learning paradigm. Additionally, comprehensive experiments conducted on the Derm7pt dataset and simulated datasets demonstrate that our method achieves superior performance while exhibiting remarkable advantages such as noise robustness, computational efficiency, and scalability. Finally, it is worth mentioning that the Joint Generalized Cosine Similarity proposed by us can not only be applied in contrastive learning, but also be easily extended to other domains.
\end{abstract}


\section{INTRODUCTION}
\begin{figure*}[htbp] % [h] 固定位置，可替换为 [htb] 等参数
\centering % 图片居中
\includegraphics[width=1\textwidth]{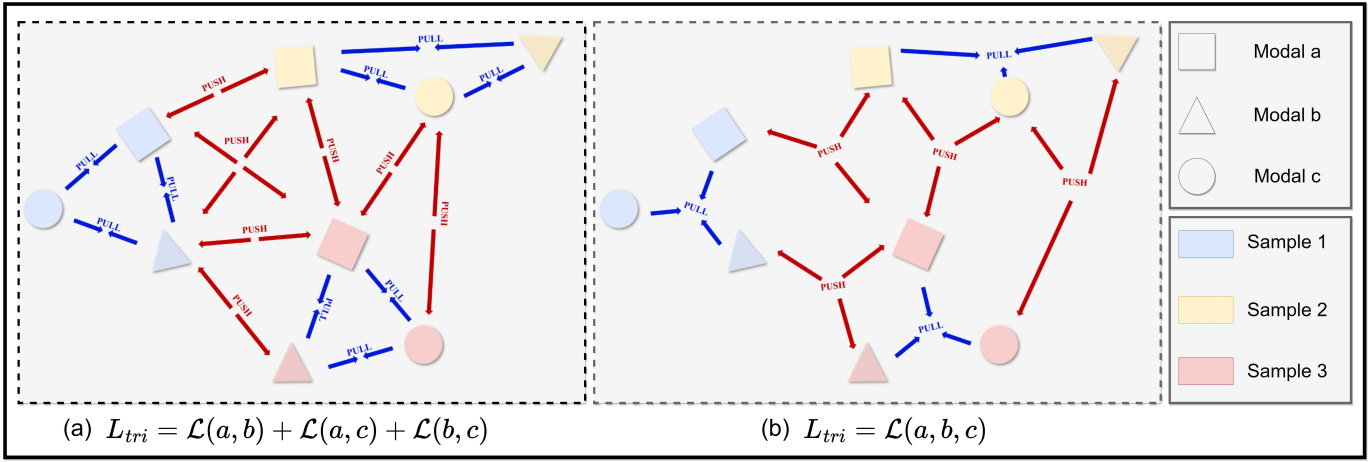}
\caption{A comparison of the adjustment of the inter-modal distances in the embedding space between the GHA Loss and the traditional pairwise-calculated loss functions.} % 图表标题
\label{contrastive} % 引用标签
\end{figure*}
In recent years, multi-modal deep learning has attracted significant attention and has been widely applied in diverse domains. Representative applications include multi-modal sentiment analysis (MSA)\cite{li2024multi, gong2024image}, audio-visual speech recognition (AVSR)\cite{li2024unified, wang2024catnet}, and medical analysis\cite{zhan2024medm2g, xu2024comprehensive, bai2024m3d}, among others. 
The primary objective of multimodal deep learning is to construct a model capable of processing information from diverse modalities and establishing associations among them. It maps the inputs from different data modalities into a shared representation space\cite{jabeen2023review}. The information derived from various modalities is contextually relevant and can provide additional complementary information to one another.
\par Multi-modal deep learning confronts various challenges. One of the directions that has attracted significant attention is the issue of multi-modal semantic alignment. In other words, it concerns how to enable the feature representations of different modalities extracted by the model to be closely mapped within the embedding space. Contrastive learning has been successfully applied to the problem of multimodal semantic alignment and has demonstrated outstanding performance\cite{zareapoor2025bimac, dufumier2024align, wang2024variance, song2024contrastalign}.
The advancement of multimodal research has led to the proliferation of complex multimodal datasets.Notable examples include the 5-modal datasets M5Product\cite{dong2022m5product} and MM-Fi\cite{yang2023mm}, the 8-modal dataset HuMMan\cite{cai2022humman}, the 10-modal dataset MedTrinity-25M\cite{xie2024medtrinity}, and the 12-modal dataset MMEarth\cite{nedungadi2024mmearth}. While modality diversity has surged, existing methods remain constrained to calculating contrastive loss in a pairwise manner and subsequently aggregating these losses into a composite loss for model training.This approach introduces three critical limitations: i)  information redundancy and conflicting optimization directions. ii) Low computational efficiency, with complexity scaling exponentially as modalities increase. iii) failure to capture joint interactions across modalities. These limitations arise from the constraints of conventional similarity measurement methods.
\par To address these challenges, we introduce a novel similarity measurement method for arbitrary numbers of vectors and propose a contrastive learning framework leveraging this method, as illustrated in \autoref{contrastive}. Our approach achieves more efficient minimization of inter-modal distances while simultaneously maximizing inter-modal separation.
Our main contributions lie in:
\begin{enumerate}
	\item We propose a novel N-dimensional similarity measurement method called Joint Generalized Cosine Similarity. Unlike traditional pairwise similarity measures, such as dot product similarity and cosine similarity, this method is capable of evaluating the similarity among multiple vectors simultaneously. Moreover, it is not only applicable to contrastive learning but can also be readily extended to other domains.
	\item Based on the proposed similarity measurement method, we introduce a corresponding contrastive learning loss function, termed GHA Loss, along with an associated contrastive learning paradigm.
	\item We conducted a comprehensive series of experiments, including ablation studies on the Derm7pt dataset using ResNet-50 and ResNet-101 models in combination with four distinct feature fusion methods. In addition, verification experiments were performed on a simulated dataset to assess noise robustness, visualize feature distributions, and evaluate computational efficiency. The results consistently demonstrate the high efficiency of the proposed method.
\end{enumerate}
The structure of this paper is organized as follows. In \autoref{s2}, we review related work and identify existing research gaps. \autoref{s3} describes the multi-modal datasets used in our experiments, along with the associated data augmentation and preprocessing techniques. In \autoref{s4}, we present a comprehensive explanation of the derivation and computation of the proposed Joint Generalized Cosine Similarity (JGCS), the corresponding contrastive learning loss function (GHA Loss), and the baseline models employed. \autoref{s5} details the experiments conducted on both real-world and simulated data, followed by a discussion of the results. Finally, we conclude the paper in \autoref{s6}.

\section{RELATED WORK}\label{s2}
\subsection{Multi-modal Deep Learning}
Multi-modal deep learning has been extensively applied in a multitude of domains and, concurrently, harbors remarkable potential for growth\cite{ferreira2025gen}. The early works include VSE++\cite{faghri2017vse}, which employs encoders independently for images and texts. Eventually, a simple dot product is utilized to represent the similarity of the features of these two modalities. The Contrastive Language-Image Pre-training (CLIP)\cite{radford2021learning} combines texts and images through the approach of contrastive learning. It is capable of learning the similarities between the two modalities of images and texts and embedding the data from both images and texts into a shared semantic space, thus serving as the foundation for numerous novel multi-modal models\cite{sun2023eva, zhang2024long}. Nevertheless, this model architecture is more intricate, and heavier encoders are employed for each modality. Models such as OSCAR and ViLBERT place a greater emphasis on modality interaction. However, they still utilize heavy convolutional neural networks to extract visual features. Models such as OSCAR\cite{li2020oscar} and ViLBERT\cite{lu2019vilbert} place a greater emphasis on modality interaction. However, they still utilize heavy convolutional neural networks to extract visual features. 
ViLT\cite{kim2021vilt} focuses on the computational efficiency of the model. It eliminates the traditional convolutional visual embedder while maintaining excellent performance.
\subsection{Loss Function For Contrastive Learning}
Contrastive learning was first applied in the field of computer science and subsequently extended to the domain of natural language analysis\cite{lin2022multimodal}. There are mainly three contrastive loss functions that are widely used. Triplet loss is a commonly employed contrastive learning loss function. By comparing the distances among the anchor sample, the positive sample, and the negative sample, it motivates the model to cluster the positive samples together while keeping them distinctly apart from the negative samples\cite{deng2018triplet}. Another widely utilized one is the InfoNCE\cite{oord2018representation} based on information theory, which has been successfully applied in the MoCo\cite{he2020momentum}. In 2020, SimCLR also proposed NT-Xent (the normalized temperature-scaled cross entropy loss), and this loss function has been extensively adopted and remains in widespread use even to this day\cite{chen2020simple}.

\subsection{Multi-modal Contrastive Learning}
Alignment is frequently employed to drive the convergence of multi-source features towards a consistent representation\cite{hou2023semantic}. The core idea of contrastive learning is to achieve alignment by measuring the similarity between sample pairs within the embedding space\cite{lin2022multimodal}. However, up to now, when the number of modalities exceeds two, most of the applications of contrastive learning can only calculate the loss in pairs for every two modalities, and then integrate these losses together in the end. \cite{you2025x2ct} proposed a three-modal contrastive learning framework named X2CT-CLIP. In this framework, the three-modal contrastive loss is obtained by calculating the InfoNCE loss values pairwise between every two modalities and then performing a weighted summation. Similarly, in \cite{yang2024tri}, the contrastive losses are also calculated pairwise between modalities, and then the weighted sum of these losses is obtained. Although the inter-modal Semi-Contrastive Learning (SCL) part in \cite{mai2022hybrid} does not explicitly perform pairwise calculations, it actually takes each modality among the three modalities of every sample as an anchor. Each anchor forms two positive pairs with the other two modalities. Subsequently, the similarities of the two positive pairs corresponding to each anchor point modality are calculated respectively, and it is required that these similarities approach the modal margin. Therefore, it still falls within the category of pairwise calculations between two modalities. In \cite{deng2018triplet}, somewhat differently, the contrastive loss among the three modalities is obtained by calculating the inter-modal triplet loss pairwise and then performing a weighted summation. \cite{ruan2024tricolo} applies the symmetric NT-Xent contrastive loss from ConVIRT\cite{zhang2022contrastive}, calculating the contrastive losses pairwise and then directly adding them together without weighting. Differing from all the above, \cite{choi2022triangular} proposes a three-modal joint contrastive loss based on the area of a triangle. Although it breaks away from the paradigm of pairwise calculations, this method has rather strong limitations and is difficult to be extended to tasks involving a larger number of modalities. Therefore, there still exists a significant research gap in developing a more universal and efficient similarity measurement method as well as a multi-modal contrastive loss function.

\begin{table*}[htbp]
	\centering
	\caption{Model Performances}
	\label{c1}
	\resizebox{\textwidth}{!}{%
		\small % 调整字号
		\renewcommand{\arraystretch}{1.2}
		\begin{tabular}{*{16}{c}} % 正确定义16列
			\Xhline{1.5pt}
			\midrule
			\multicolumn{2}{c}{} & Acc & $\kappa$ & P & R & $F_1$ & AUC & 
			\multicolumn{2}{c}{} & Acc & $\kappa$ & P & R & $F_1$ & AUC \\
			\midrule
			
			% 第1行（验证列数）
			\multirow{3}{*}{Res50+Concat} & Normal & 58.07 & 0.4423 & 37.87 & 33.03 & 0.3346 & 0.8329 & 
			\multirow{3}{*}{Res101+Concat} & Normal & 59.85 & 0.4626 & 40.85 & 31.70 & 0.3085 & 0.8487 \\ % 共15个&
			\cmidrule(lr){2-8} \cmidrule(lr){10-16}
			& Dual & 60.94 & 0.4750 & $\mathbf{47.69}$ & 30.94 & 0.3407 & 0.9005 &
			& Dual & 61.72 & 0.4917 & $\mathbf{47.98}$ & $\mathbf{37.85}$ & 0.3765 & $\mathbf{0.9044}$ \\
			\cmidrule(lr){2-8} \cmidrule(lr){10-16}
			& GHA & $\mathbf{62.76}$ & $\mathbf{0.5111}$ & 39.25 & $\mathbf{45.41}$ & $\mathbf{0.3934}$ & $\mathbf{0.9046}$ &
			& GHA & $\mathbf{63.80}$ & $\mathbf{0.5226}$ & 47.26 & 37.77 & $\mathbf{0.3973}$ & 0.8675 \\
			\midrule
			
			\multicolumn{2}{c}{} & Acc & $\kappa$ & P & R & $F_1$ & AUC & 
			\multicolumn{2}{c}{} & Acc & $\kappa$ & P & R & $F_1$ & AUC \\
			\midrule
			
			% 第1行（验证列数）
			\multirow{3}{*}{Res50+Sum} & Normal & 53.39 & 0.3951 & 27.53 & 31.08 & 0.2804 & 0.7916 & 
			\multirow{3}{*}{Res101+Sum} & Normal & 45.31 & 0.2833 & 29.36 & 26.81 & 0.2551 & 0.8113 \\ % 共15个&
			\cmidrule(lr){2-8} \cmidrule(lr){10-16}
			& Dual & 57.02 & 0.4262 & $\mathbf{38.82}$ & 31.74 & 0.3240 & 0.8325 &
			& Dual & 52.08 & 0.3587 & $\mathbf{34.98}$ & 28.22 & 0.2838 & 0.7785 \\
			\cmidrule(lr){2-8} \cmidrule(lr){10-16}
			& GHA & $\mathbf{60.42}$ & $\mathbf{0.4776}$ & 37.80 & $\mathbf{38.91}$ & $\mathbf{0.3601}$ & $\mathbf{0.8583}$ &
			& GHA & $\mathbf{55.21}$ & $\mathbf{0.4133}$ & 32.41 & $\mathbf{35.44}$ & $\mathbf{0.3255}$ & $\mathbf{0.8380}$ \\
			\midrule
			
			\multicolumn{2}{c}{} & Acc & $\kappa$ & P & R & $F_1$ & AUC & 
			\multicolumn{2}{c}{} & Acc & $\kappa$ & P & R & $F_1$ & AUC \\
			\midrule
			
			% 第1行（验证列数）
			\multirow{3}{*}{Res50+Attn} & Normal & 57.03 & 0.4271 & 38.61 & 29.14 & 0.3030 & 0.7855 & 
			\multirow{3}{*}{Res101+Attn} & Normal & 59.38 & 0.4448 & 33.77 & 28.72 & 0.2920 & 0.8437 \\ % 共15个&
			\cmidrule(lr){2-8} \cmidrule(lr){10-16}
			& Dual & 59.38 & 0.4490 & 36.31 & 32.25 & 0.3259 & 0.8309 &
			& Dual & $\mathbf{61.46}$ & 0.4701 & $\mathbf{44.35}$ & $\mathbf{33.86}$ & $\mathbf{0.3515}$ & 0.8749 \\
			\cmidrule(lr){2-8} \cmidrule(lr){10-16}
			& GHA & $\mathbf{63.02}$ & $\mathbf{0.4922}$ & $\mathbf{49.29}$ & $\mathbf{35.04}$ & $\mathbf{0.3782}$ & $\mathbf{0.8639}$ &
			& GHA & 60.94 & $\mathbf{0.4817}$ & 40.72 & 33.72 & 0.3413 & $\mathbf{0.8889}$ \\
			\midrule
			
			\multicolumn{2}{c}{} & Acc & $\kappa$ & P & R & $F_1$ & AUC & 
			\multicolumn{2}{c}{} & Acc & $\kappa$ & P & R & $F_1$ & AUC \\
			\midrule
			
			% 第1行（验证列数）
			\multirow{3}{*}{Res50+Gate} & Normal & 54.43 & 0.4176 & 38.78 & 35.13 & 0.3098 & 0.8236 & 
			\multirow{3}{*}{Res101+Gate} & Normal & 54.69 & 0.3951 & 32.77 & 27.30 & 0.2871 & 0.7917 \\ % 共15个&
			\cmidrule(lr){2-8} \cmidrule(lr){10-16}
			& Dual & $\mathbf{65.89}$ & $\mathbf{0.5394}$ & 42.14 & 36.20 & 0.3655 & $\mathbf{0.8964}$ &
			& Dual & $\mathbf{63.02}$ & $\mathbf{0.4959}$ & 36.75 & 31.20 & 0.3234 & $\mathbf{0.8940}$ \\
			\cmidrule(lr){2-8} \cmidrule(lr){10-16}
			& GHA & 63.54 & 0.5165 & $\mathbf{44.51}$ & $\mathbf{39.56}$ & $\mathbf{0.4032}$ & 0.8817 &
			& GHA & 59.90 & 0.4819 & $\mathbf{42.43}$ & $\mathbf{39.85}$ & $\mathbf{0.3861}$ & 0.8405 \\
			\midrule
		\end{tabular}
	}
\end{table*}

\begin{figure}[htbp] % [h] 固定位置，可替换为 [htb] 等参数
\centering % 图片居中
\includegraphics[width=0.4\textwidth]{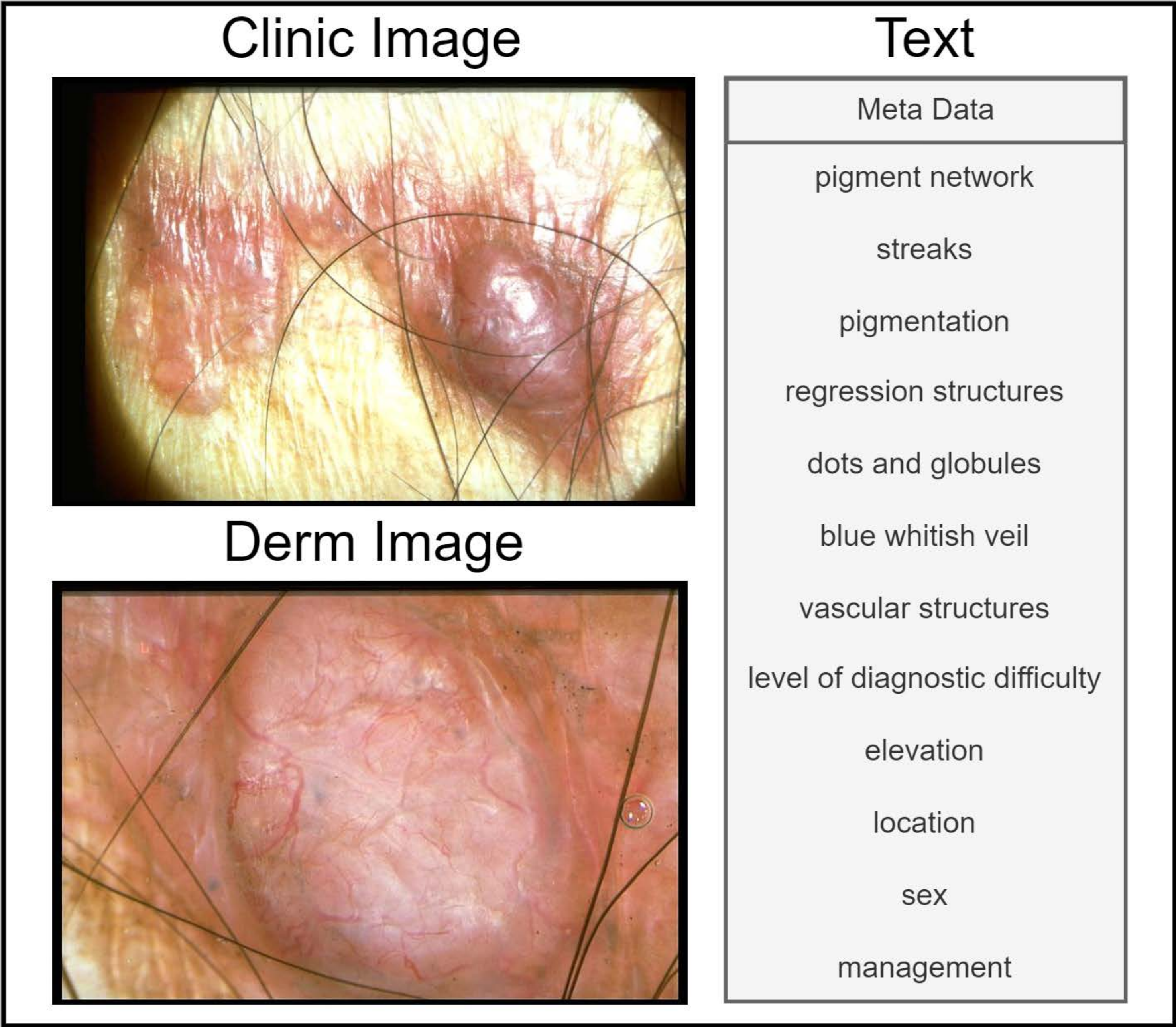}
\caption{An example of the three modalities in the Derm7pt dataset.} % 图表标题
\label{derm7pt} % 引用标签
\end{figure}

\begin{figure*}[htbp] % [h] 固定位置，可替换为 [htb] 等参数
\centering % 图片居中
\includegraphics[width=0.9\textwidth]{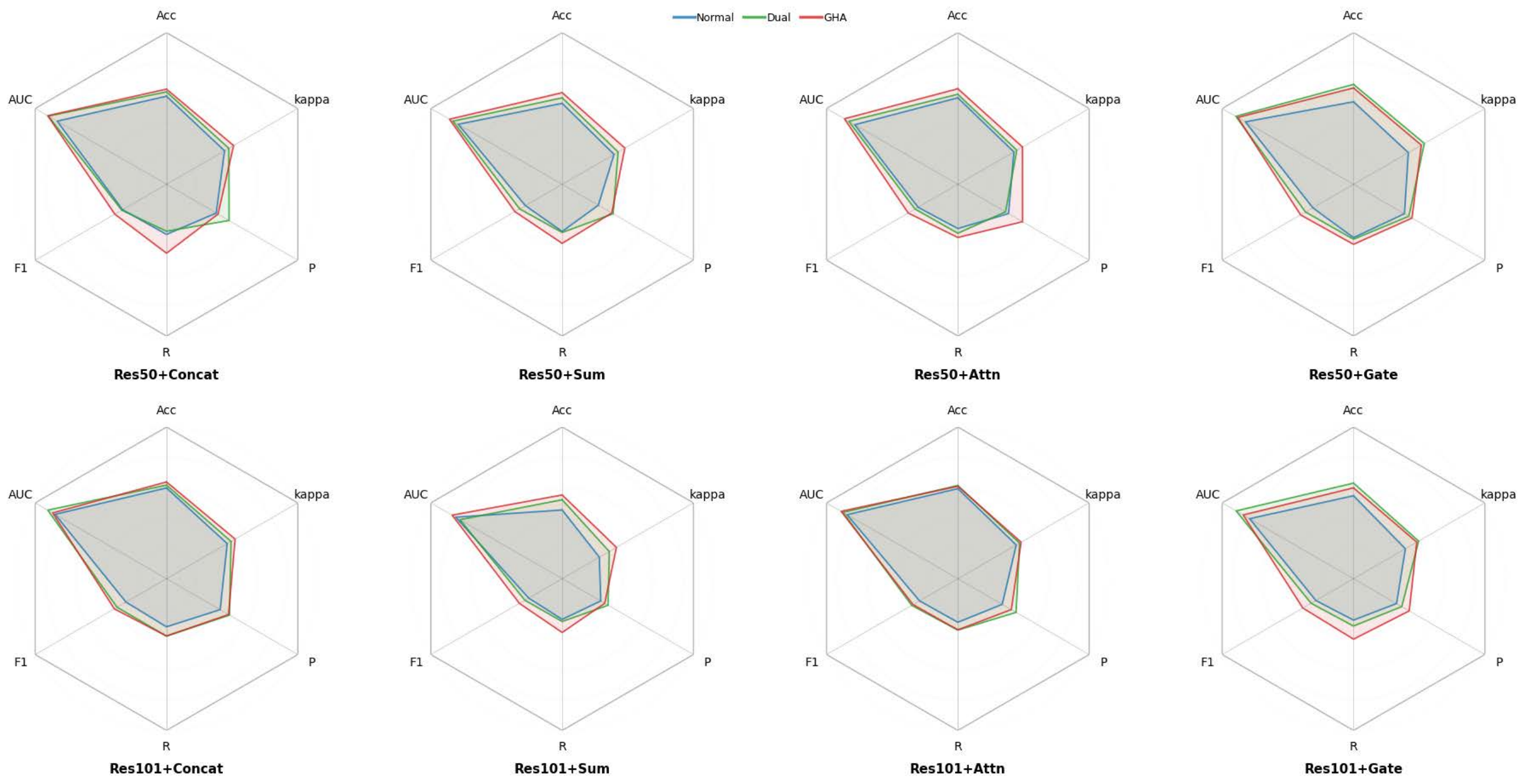}
\caption{The radar visualization graphs of eight baseline models in terms of six evaluation indicators.} % 图表标题
\label{radar} % 引用标签
\end{figure*}

\begin{figure*}[htbp] % [h] 固定位置，可替换为 [htb] 等参数
\centering % 图片居中
\includegraphics[width=1\textwidth]{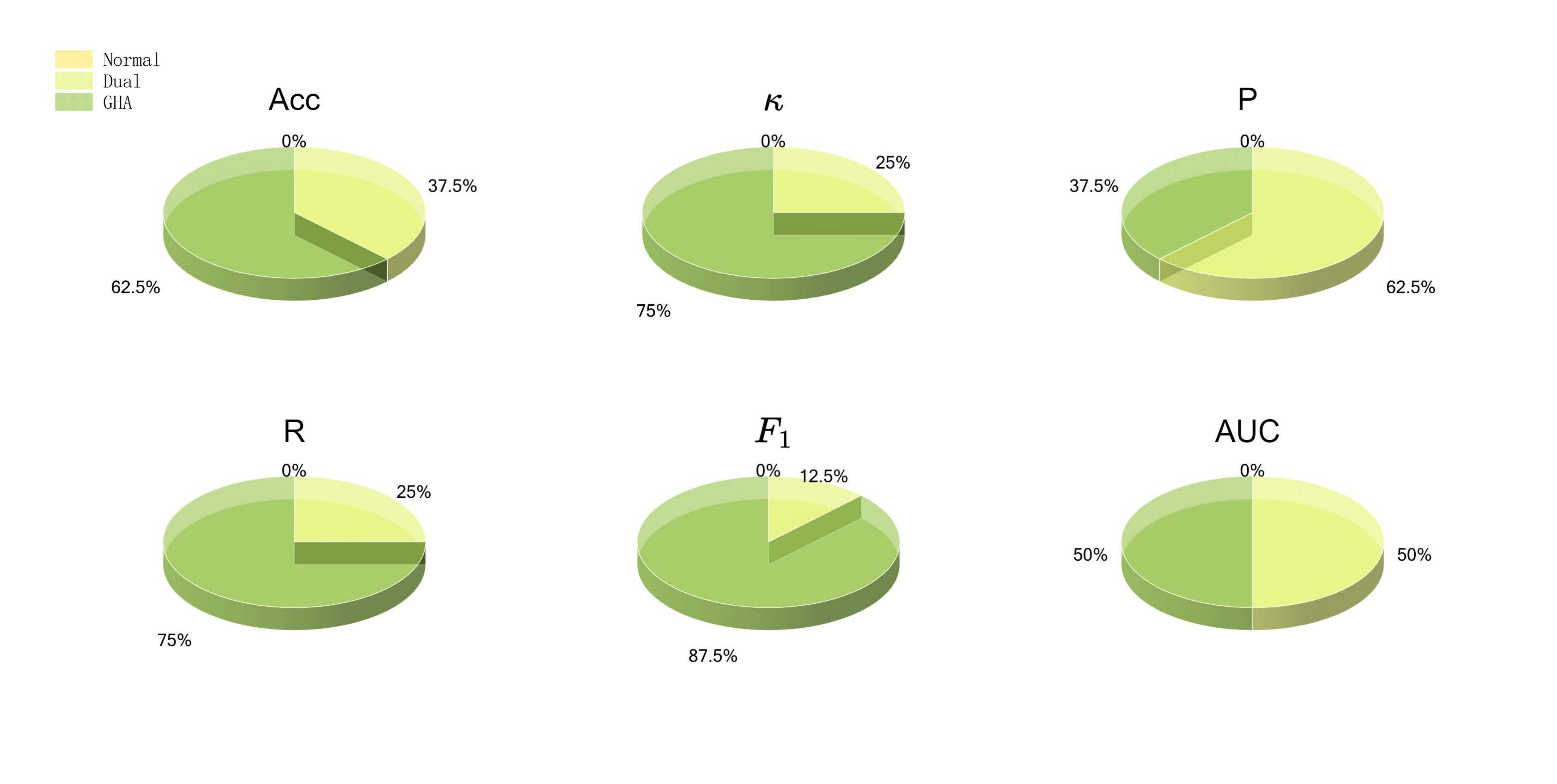}
\caption{The number of times each of the three methods outperforms others in terms of the six indicators.} % 图表标题
\label{pie} % 引用标签
\end{figure*}
\begin{figure*}[htbp] % [h] 固定位置，可替换为 [htb] 等参数
	\centering % 图片居中
	\includegraphics[width=0.8\textwidth]{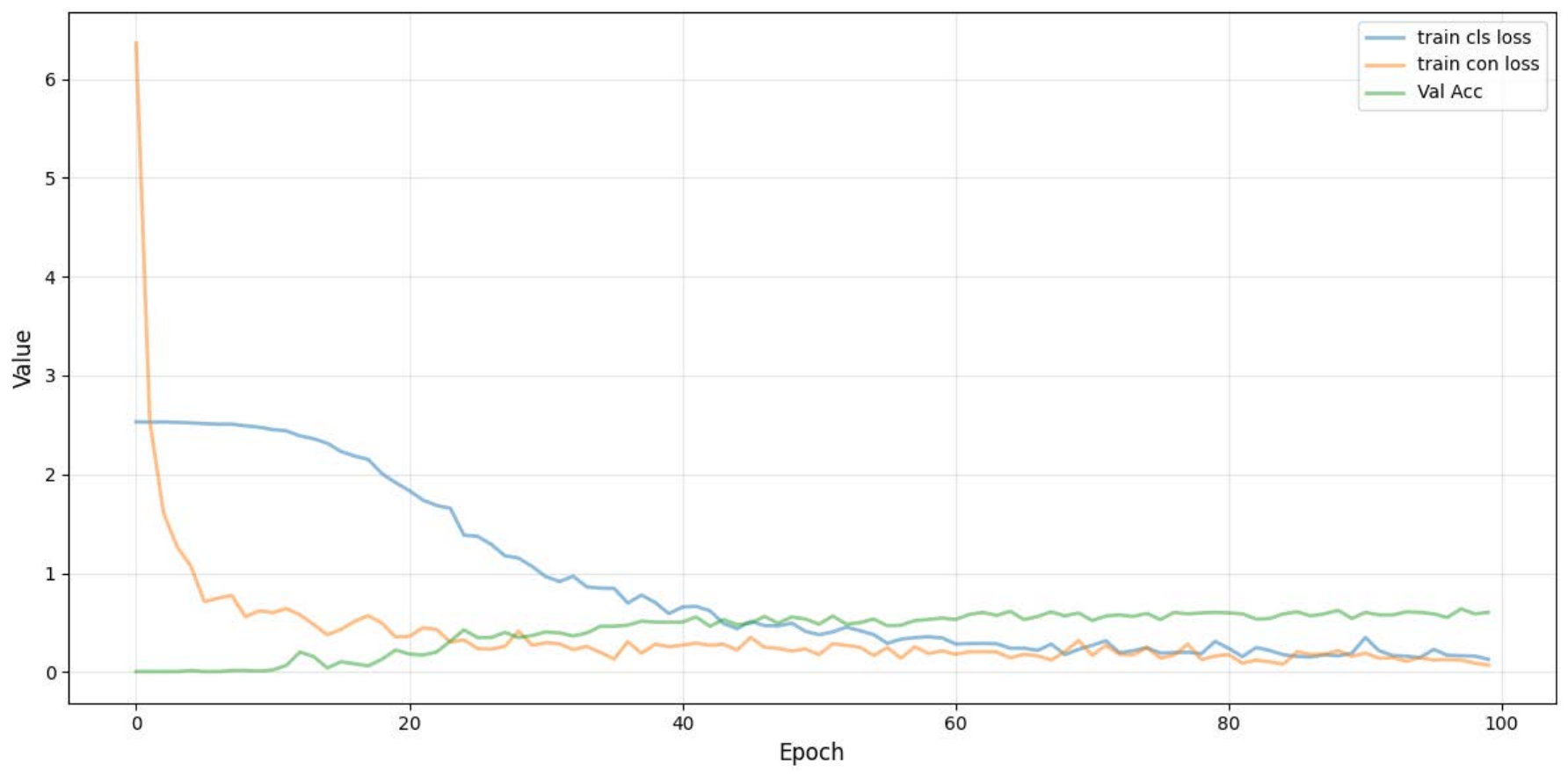}
	\caption{
		The convergence performance of the Res101+Concat model. The train cls loss represents the classification loss value calculated by the Focal loss, the train con loss represents the contrastive loss value calculated by the GHA loss, and the Val Acc represents the Top-1 accuracy rate on Validation dataset.} % 图表标题
	\label{converge} % 引用标签
\end{figure*}

\begin{figure*}[htbp] % [h] 固定位置，可替换为 [htb] 等参数
\centering % 图片居中
\includegraphics[width=0.9\textwidth]{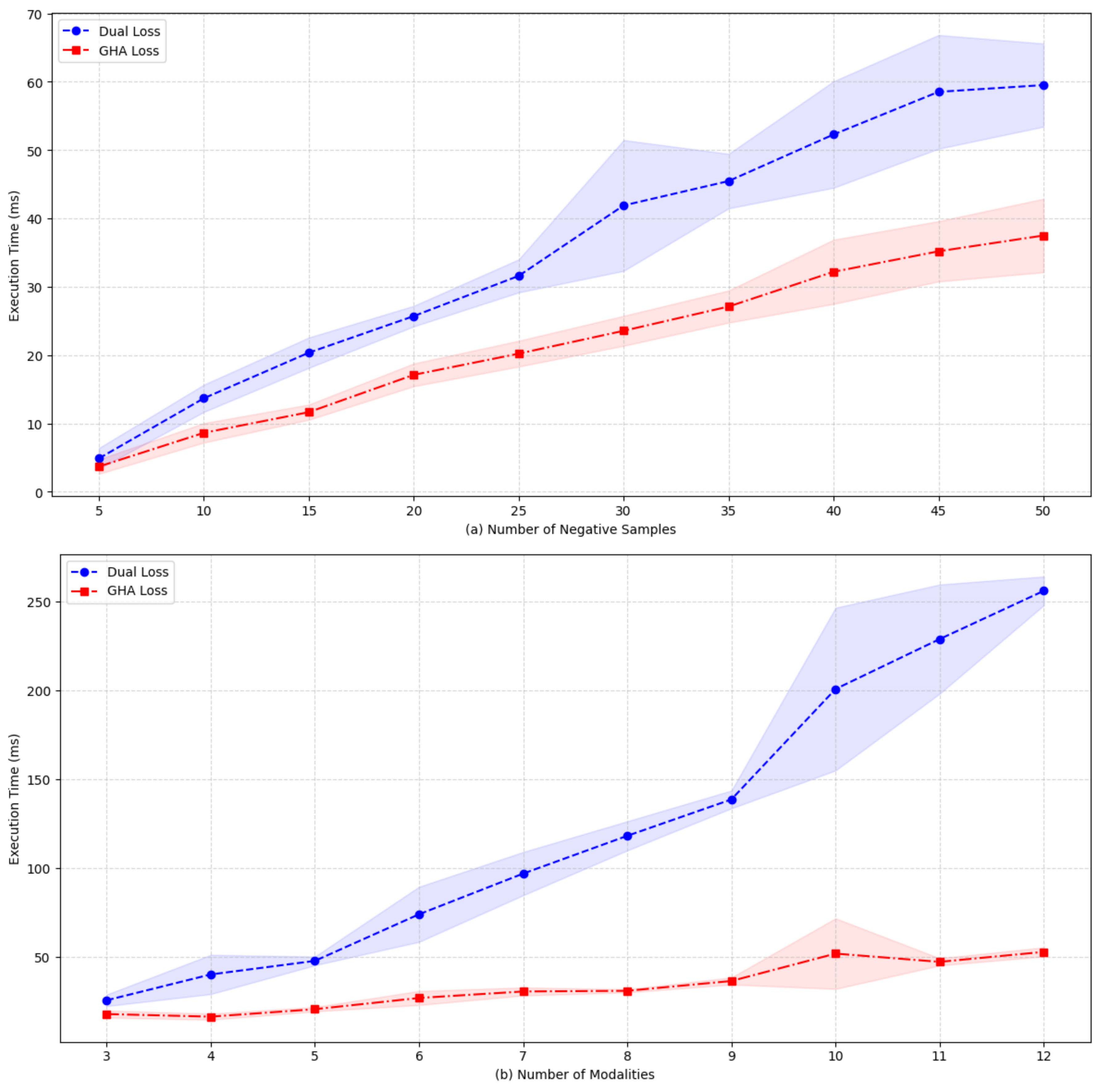}
\caption{A comparison of the computational efficiency of GHA Loss and Dual Loss under different numbers of negative samples and different numbers of modalities.} % 图表标题
\label{runtime_compare} % 引用标签
\end{figure*}

\section{DATASET}\label{s3}

The dataset used in this study is Derm7pt \cite{kawahara2018seven}, a three-modal dataset designed for skin cancer diagnosis. An example of a dataset instance is shown in \autoref{derm7pt}. Derm7pt contains 1,011 multi-modal instances, each comprising three modalities: clinical images, dermatoscopic images, and metadata in textual form. The dataset is officially partitioned into a training set with 413 instances, a validation set with 203 instances, and a test set with 395 instances, covering a total of 15 label categories.
\par For the clinical images, horizontal flipping and color perturbation were applied as data augmentation techniques. For the dermatoscopic images, vertical flipping, rotation, and sharpening augmentations were employed. Both types of images were resized to a uniform resolution of $224 \times 224$, and channel normalization was performed to ensure a mean of 0 and a variance of 1. As for the metadata, one-hot encoding was applied to each categorical feature.

\section{METHODS}\label{s4}
\begin{figure}[htbp] % [h] 固定位置，可替换为 [htb] 等参数
	\centering % 图片居中
	\includegraphics[width=0.4\textwidth]{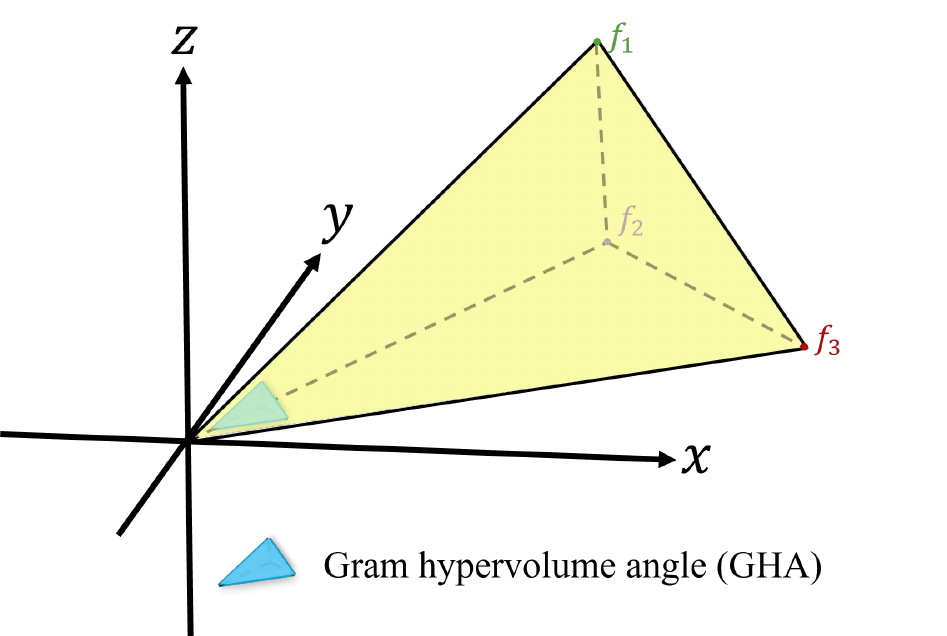}
	\caption{The visualization of the Gram hypervolume angle (GHA).} % 图表标题
	\label{gha_show} % 引用标签
\end{figure}
\subsection{Joint Generalized Cosine Similarity}\label{gha}
\par Conventional similarity measurement method, such as Euclidean similarity, dot product similarity, and cosine similarity, are inherently limited to quantifying the similarity between two vectors. When confronted with the necessity of computing similarity across more than two modal feature vectors, the typical approach involves performing individual similarity computations for each pair of modal vectors, followed by a weighted summation to yield an aggregate similarity score. 
However, these conventional approaches are beset by several shortcomings: i) They are incapable of modeling the collective interaction among multi-modal feature vectors, being constrained to capture only pairwise modal interactions. ii) They exhibit vulnerabilities to noise sensitivity and suffer from information redundancy.
Therefore, to address these issues, we propose a method capable of computing the joint similarity between multi-modal feature vectors in a single step.
\par 
For a matrix $\mathbf{M}=[\mathbf{f_1}, \mathbf{f_2}, ... , \mathbf{f_n}]^\text{T}$, wherein each row vector $\mathbf{f_i}^\text{T} \in \mathbb{R}^{1 \times D}$ represents a feature vector extracted by an encoder from a single modality input, and $n$ denotes the number of modalities and $D$ signifies the dimensionality of the feature vectors.
The absolute value of the determinant of an n-dimensional square matrix has a geometric significance, which represents the hypervolume of the n-dimensional solid spanned by its row (or column) vectors. 
For this non-square matrix $\mathbf{M}$, we can derive the hypervolume $V$ of the n-dimensional space spanned by these n feature vectors by calculating the determinant of the Gram matrix constructed from its row vectors:
\begin{equation}
	V = |\text{det}(\mathbf{M})| = \sqrt{\text{det}(\mathbf{M}\mathbf{M}^T)}
\end{equation}
Subsequently, in order to maintain conciseness while still retaining generality, we utilize the example of three-modal feature vector to conduct the derivations presented in the subsequent content. For the three feature vectors ${\mathbf{f}, \mathbf{g}, \mathbf{k}}$ , their Gram matrix $\mathbf{G}$ can be expressed as:
\begin{equation}
	\mathbf{G} = \begin{bmatrix} % 括号为方括号
		\mathbf{f}\cdot\mathbf{f} & \mathbf{f}\cdot\mathbf{g} &\mathbf{f}\cdot\mathbf{k} \\
		\mathbf{g}\cdot\mathbf{f} & \mathbf{g}\cdot\mathbf{g} &\mathbf{g}\cdot\mathbf{k} \\
		\mathbf{k}\cdot\mathbf{f} & \mathbf{k}\cdot\mathbf{g} &\mathbf{k}\cdot\mathbf{k} \\
	\end{bmatrix}
\end{equation}
By calculating its determinant, we can obtain the following representation:
\begin{equation}
	\text{det}(\mathbf{G}) = \|\mathbf{f}\|_2^2\|\mathbf{g}\|_2^2\|\mathbf{k}\|_2^2\cdot[1-\Phi(\mathbf{f},\mathbf{g},\mathbf{k})]
\end{equation}
\begin{equation}
	\Phi(\mathbf{f},\mathbf{g},\mathbf{k}) = \text{cos}^2\theta_{fg}+\text{cos}^2\theta_{fk}+\text{cos}^2\theta_{gk}-2\text{cos}\theta_{fg}\text{cos}\theta_{fk}\text{cos}\theta_{gk}
\end{equation}
Among them, for $\Phi(\mathbf{f},\mathbf{g},\mathbf{k})$ it satisfies certain characteristics: i) Due to the non-negativity of the determinant of the Gram matrix $\mathbf{G}$, the value of $\Phi(\mathbf{f},\mathbf{g},\mathbf{k})$ is less than or equal to 1. ii)When the three vectors are orthogonal, the determinant value of the Gram matrix $\mathbf{G}$ reaches its maximum, and at this time, the value of $\Phi(\mathbf{f},\mathbf{g},\mathbf{k})$ has a minimum value of 0. Therefore, we can regard the function $\Phi(\mathbf{f},\mathbf{g},\mathbf{k})$ as a special cosine function $\text{cos}^2\Theta_{\mathbf{f},\mathbf{g},\mathbf{k}}$ (i.e. $\text{cos}^2\Theta_{\mathbf{f},\mathbf{g},\mathbf{k}} \triangleq \Phi(\mathbf{f},\mathbf{g},\mathbf{k}) $), and $\Theta$ is named the Gram hypervolume angle (GHA), represents a generalized angle derived from the determinant of the Gram matrix. We visualized GHA, as shown in \autoref{gha_show}, in the case of trimodality, it can be visualized as a small vertebral body at the tip of a triangular pyramid.

\par For the matrix $\mathbf{M}$ composed of n modals mentioned above, we have the general calculation formula of the Gram hypervolume angle (GHA):
\begin{equation}\label{theta}
	\Theta_{\mathbf{f_1},\mathbf{f_2},...,\mathbf{f_n}}  = \text{arcsin}(\frac{\sqrt{\text{det}(\mathbf{M}\mathbf{M}^T)}}{\prod_{i=1}^{n} \|\mathbf{f}_i\|_2})
\end{equation}
Finally, in accordance with the general similarity measurement approach, we propose using $\text{cos}\Theta_{\mathbf{f_1},\mathbf{f_2},...,\mathbf{f_n}}$ (abbreviated as $\text{cos}\Theta$ in the following text) as the practical calculation method for similarity measurement.
This method of similarity measurement has been named by us as the joint generalized cosine similarity.
\par The GHA $\Theta$ derived from the Gram determinant exhibits exceptional physical significance and mathematical properties, as listed below. 
\begin{enumerate}
	\item \textbf{Downward Compatibility}: When the number of modal feature vector is two, the GHA degenerates into the conventional 2D planar angle.
	\item \textbf{Coordinate Rotation Invariance}: The GHA remains invariant under orthogonal transformations (rotations/reflections) applied to the feature vectors, as shown in \autoref{proof1}.
	\begin{equation}\label{proof1}
		\mathbf{G}'_{ij} = \mathbf{f}'_i \cdot \mathbf{f}'_j = (\mathbf{Q}\mathbf{f}_i)^T(\mathbf{Q}\mathbf{f}_j) = \mathbf{f}_i^T \mathbf{Q}^T \mathbf{Q} \mathbf{f}_j = \mathbf{f}_i \cdot \mathbf{f}_j
	\end{equation}
	where $\mathbf{Q} \in \mathbb{R}^{D \times D}$ is an orthogonal matrix, and $\mathbf{f}'_i = \mathbf{Q}\mathbf{f}_i$ is the transformed feature vector. Therefore, $\det(\mathbf{G}') = \det(\mathbf{G})$, and $\|\mathbf{f}'_i\|_2 = \|\mathbf{f}_i\|_2$, the GHA $\Theta$ remains unchanged.
	\item \textbf{Permutation Symmetry}: Swapping the order of feature vectors only alters the sign of the Gram matrix determinant while keeping its absolute value unchanged, thus leaving the GHA $\Theta$ unaffected.
	\item \textbf{Extremal Properties}: If the vectors are linearly correlated, then $\det(\mathbf{G}) = 0$, hence $\Theta = 0$, $\cos \Theta = 1$, similarity is maximum. While, if the vectors are pairwise orthogonal and $n \leq D$, the Gram matrix is diagonal, $\det(\mathbf{G}) = \prod \|\mathbf{f}_i\|^2_2$, at this time $\Theta = \frac{\pi}{2}$, $\cos \Theta = 0$, similarity is minimum. 
\end{enumerate}
In summary, the GHA method inherits the intuitive properties of traditional cosine similarity and generalizes them to high-dimensional spaces. It is built on a solid mathematical foundation and exhibits invariance, making it well-suited for measuring the similarity of multi-modal features.
\par Finally, it is important to note that since GHA is derived by squaring and then taking the square root, the calculation of this method is symmetric in both the positive and negative directions. In other words, the GHA value calculated from the three feature vectors $\{a,b,c\}$ is identical to the GHA value calculated from the three feature vectors $\{d,b,c\}$ where $d = - a$. To address this symmetry, we must impose a constraint on the modality to ensure that the values of all feature vectors are positive. This can be easily achieved by adding a ReLU activation function to the final layer of the model.

\subsection{Multi-Vector Joint Contrastive Learning Loss Function}
The similarity measurement method we proposed can be diversely applied in contrastive learning. Here, taking the InfoNCE loss as the framework, we construct a new paradigm for the contrastive learning loss function $\mathcal{L}_{\mathcal{C}}$:
\begin{equation}
	\mathcal{L}_{\mathcal{C}} = -\frac{1}{B} \sum_{i=1}^B \log \frac{
		\exp\left( \frac{\cos\Theta_{\text{pos}}^{(i)}}{ \tau } \right)
	}{
		\exp\left( \frac{\cos\Theta_{\text{pos}}^{(i)}}{ \tau } \right) + 
		\sum_{j=1}^{K} \exp\left( \frac{\cos\Theta_{\text{neg},j}^{(i)}}{ \tau } \right)
	}
\end{equation}
Among them, $B$ represents the batch size, $\cos\Theta_{\text{pos}}^{(i)}$ represents the joint generalized cosine similarity of the i-th positive n-tuple, $\cos\Theta_{\text{neg},j}^{(i)}$ represents the joint generalized cosine similarity of the j-th negative n-tuple of the i-th sample.$\tau$ is an adjustable temperature parameter. $K$ represents the number of negative n-tuple combinations generated for each original sample.
It is worth noting that the definition of positive and negative sample pairs is determined by the task. In the experiment of \autoref{s5}, we employed a simple inter-modal method. That is to say, three modalities in each existing sample form a group of positive samples, while three modalities groups randomly selected constituted the negative samples.
\par Meanwhile, in order to enable multiple modalities to be evenly drawn closer in terms of similarity, and to avoid the situation where the feature vectors of any two modalities are pulled onto the same plane, thus preventing the feature vectors of the remaining modalities from being further optimized, we propose the angular equilibrium loss function $\mathcal{L}_{\mathcal{A}}$:
\begin{equation}
	\mathcal{L}_{\mathcal{A}} = \frac{1}{B} \sum_{i=1}^B \left[ 
	\frac{1}{\binom{n}{2}} \sum_{1 \leq m < k \leq n} \left( c_{i}^{(m,k)} - \bar{c}_i \right)^2 
	\right]
\end{equation}
Here, $n$ represents the number of modalities.  $c_{i}^{(m,k)}$ is the cosine similarity between the m-th and the k-th modalities of the i-th sample, which can be readily obtained through the off-diagonal elements of the Gram matrix. $\bar{c}_i$ is the average cosine similarity of all modality pairs of the i-th sample, and $B$ is the batch size. This constraint serves as a regularization constraint, restricting the variance of the cosine similarity of each modality pair. It prevents certain modalities from being either overly similar or overly dissimilar, encourages the model to make equal use of information from all modalities, and mitigates the problem of overfitting.
\par 
Finally, we obtain the complete loss function, named the GHA loss, which is employed for the semantic alignment among the modal feature vectors:
\begin{equation}
	\mathcal{L}_{\text{GHA}} = \mathcal{L}_{\mathcal{C}} + \lambda \cdot \mathcal{L}_{\mathcal{A}}
\end{equation}
Where, $\lambda$ is an adjustable regularization coefficient

\subsection{Model}
In this section, we constructed eight distinct baseline models through the combination of two disparate image encoders (ResNet-50, ResNet-101) and four unique feature fusion methodologies, as illustrated in Figure 1. These models were established to facilitate subsequent validation of the efficacy of our proposed GHA loss function. 
\par For inputs of three modalities: i) The dermoscopic images $\mathbf{D} \in \mathbb{R}^{C \times H \times W}$ ii) The clinical images $\mathbf{C} \in \mathbb{R}^{C \times H \times W}$ iii) Metadata text $\mathbf{m} \in \mathbb{R}^{m}$
A three-branch encoder is employed to perform feature extraction on them respectively:
\begin{align}
	\mathbf{f}_d &= \sigma(\mathbf{W}_d \cdot \mathcal{F}_{\text{Encoder}}(\mathbf{D}) + \mathbf{b}_d) \in \mathbb{R}^{256} \\
	\mathbf{f}_c &= \sigma(\mathbf{W}_c \cdot \mathcal{F}_{\text{Encoder}}(\mathbf{C}) + \mathbf{b}_c) \in \mathbb{R}^{256} \\
	\mathbf{f}_t &= \sigma(\mathbf{W}_{m2} \cdot \sigma(\mathbf{W}_{m1} \mathbf{m} + \mathbf{b}_{m1}) + \mathbf{b}_{m2}) \in \mathbb{R}^{256}
\end{align}
Among them, $\mathcal{F}_{\text{Encoder}}$ denotes an instance of the two encoders mentioned above, $\sigma(\cdot)$represents the ReLU activation function (Corresponding to the constraint mentioned in \autoref{gha}), and $\mathbf{W}$ and $\mathbf{b}$ are learnable parameters. Here, the feature vector of each modality is unified to a length of 256.
After that, The four feature fusion methodologies are respectively specified by the following equations:
\begin{enumerate}
	\item \textbf{Concat}: Simply concatenating the three feature vectors in order, as shown in \autoref{concat}.
	\begin{equation}\label{concat}
		\mathbf{f}_{\text{fused}} = \begin{bmatrix} % 括号为方括号
			\mathbf{f}_d \\
			\mathbf{f}_c \\
			\mathbf{f}_t \\
		\end{bmatrix} \in \mathbb{R}^{768}
	\end{equation}
	This method can effectively preserve the information in the original feature vectors. However, it runs the risk of having an excessively high dimensionality.
	
	\item \textbf{Weighted Sum}: This method introduces learnable modality weight parameters \(\omega = [\omega_d, \omega_c, \omega_m]^T \in \mathbb{R}^3\) that are normalized by the sigmoid function, and performs a weighted sum of the three feature vectors, as shown in \autoref{sum}.
	\begin{equation}\label{sum}
		\mathbf{f}_{\text{fused}} = \omega_d \odot \mathbf{f}_d + \omega_c \odot \mathbf{f}_c + \omega_m \odot \mathbf{f}_m \in \mathbb{R}^{256}
	\end{equation}
	Where \(\odot\) denotes the Hadamard product (element-wise multiplication). This approach is computationally simple and has a small number of parameters but fails to capture complex interactions between modalities.
	
	\item \textbf{Multi-head Attention Mechanism}: 
	This method utilizes residual connection and the multi-head attention mechanism for the dynamic fusion of feature vectors. First, a learnable embedding matrix \(\mathbf{E} \in \mathbb{R}^{3 \times 256}\) is used to obtain the modality embedding \(\mathbf{e} \in \mathbb{R}^{256}\) of the input feature vector, and then it is added to the original feature vector through a residual connection:
	\begin{equation}
	\mathbf{H}_0 = \begin{bmatrix}
		\mathbf{f}_d + \mathbf{e}_d \\
		\mathbf{f}_c + \mathbf{e}_c \\
		\mathbf{f}_m + \mathbf{e}_m
	\end{bmatrix} \in \mathbb{R}^{3 \times 256}
	\end{equation}
	Here, the modality embeddings $\mathbf{e}$ are directly generated through modality IDs (0, 1, 2). The residual connection here is equivalent to adding a unique embedding vector to the features of each modality, enabling the model to distinguish among dermoscopy, clinical image, and text features.
	Next, based on the multi-head attention mechanism $\text{Attention}(\cdot)$, after calculating the outputs of the same dimension, perform residual connection and the non-linear transformation:
	\begin{align}
		\mathbf{H}_1 &= \mathbf{H}_0 + \text{Attention}(\mathbf{Q},\mathbf{K},\mathbf{V}) \in \mathbb{R}^{3 \times 256} \\
		\mathbf{H}_2 &= \text{GELU}(\mathbf{H}_1 \mathbf{W}_f^{(1)} + \mathbf{b}_f^{(1)}) \in \mathbb{R}^{3 \times 256} \\
		\mathbf{H}_3 &= \mathbf{H}_2 \mathbf{W}_f^{(2)} + \mathbf{b}_f^{(2)} \in \mathbb{R}^{3 \times 256}
	\end{align}
	Where $\mathbf{Q},\mathbf{K},\mathbf{V}$ are all is the $\mathbf{H}_0$.
	Finally, perform average pooling on the $\mathbf{H}_3$ containing the three feature representations to obtain the final fused feature:
	\begin{equation}
		\mathbf{f}_{\text{fused}} = \frac{1}{3}\sum_{i=1}^3 \mathbf{H}_3^{[i]} \in \mathbb{R}^{256}
	\end{equation}
	Where $\mathbf{H}_3^{[i]}$ represents the i-th row of the matrix $\mathbf{H}_3$.
	This method can dynamically capture interactions among multiple modalities, but it is computationally complex.
	
	\item \textbf{Gate Mechanism}: This method weights the summation process of the three feature vectors through gate scores, as shown in \autoref{gate}.
	\begin{equation}\label{gate}
		\mathbf{f}_{\text{fused}} =  \sum_{i \in \{d,c,m\}} \underbrace{\sigma\Big( \mathbf{W}_g \begin{bmatrix} % 括号为方括号
				\mathbf{f}_d \\
				\mathbf{f}_c \\
				\mathbf{f}_m \\
			\end{bmatrix} \Big)_i}_{\text{gate scores}} \odot \mathbf{f}_i 
	\end{equation}
	Among them, $\mathbf{W}_g$ is the learnable gate weight matrix, $\sigma(\cdot)$ is the Sigmoid activation function, and $\odot$ is element-wise multiplication. 
\end{enumerate}

\section{EXPERIMENT AND RESULT}\label{s5}
\begin{figure*}[htbp] % [h] 固定位置，可替换为 [htb] 等参数
	\centering % 图片居中
	\includegraphics[width=1\textwidth]{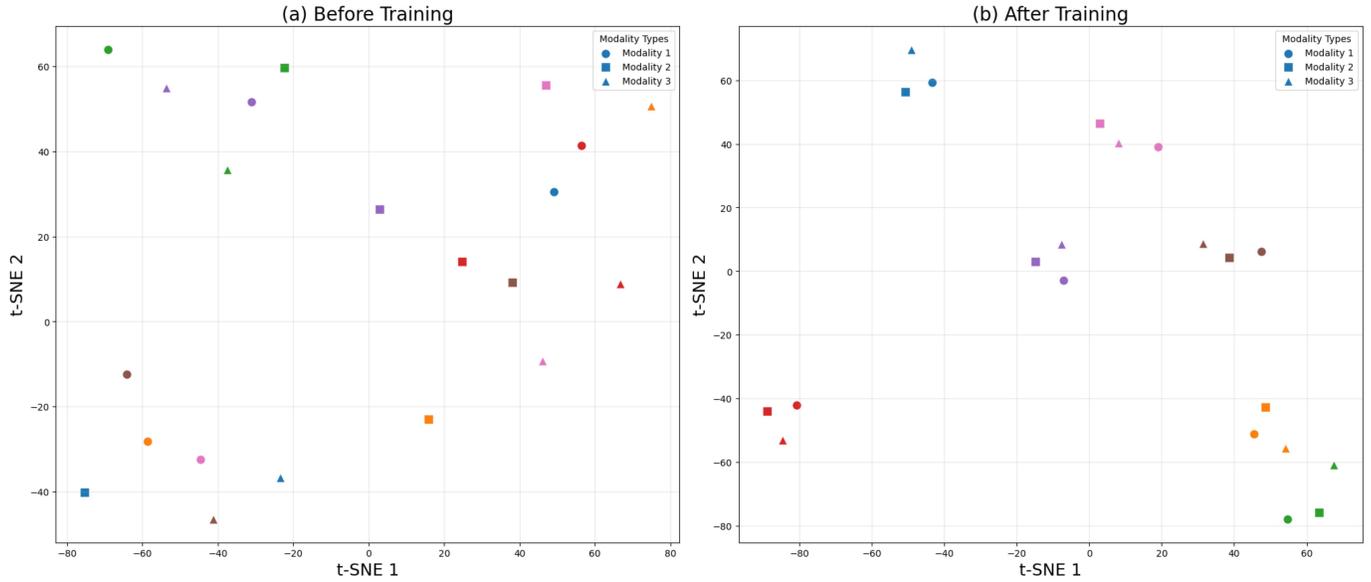}
	\caption{Two-dimensional embedding diagrams drawn by the t-SNE algorithm before and after the model uses the GHA Loss for inter-modal contrasive learning. In the diagrams, the three modalities of the same color belong to the same instance (sample).} % 图表标题
	\label{distribute} % 引用标签
\end{figure*}
\subsection{Experimental Configuration}
%设备环境
In the present study, experiments using real data were conducted on an NVIDIA GeForce RTX 4070 Laptop GPU.
%超参设置
The hyperparameters were configured as follows: the optimization was performed using the Adam algorithm \cite{kingma2014adam} with a fixed learning rate of 7e-5. The batch size was uniformly set to 24, and the number of training epochs was set to 100. For classification, the Focal Loss \cite{lin2017focal} was used as the primary loss function, without class weighting, and the gamma parameter was set to 1. The data loader employed a class-balanced sampling strategy, ensuring that all classes had an equal probability of being sampled within each mini-batch.
%比对对象和损失函数使用
For each baseline model, three distinct experimental settings were evaluated for comparison:
(i \textbf{Normal}: Without using contrastive learning.
(ii \textbf{Dual}: Incorporating a pairwise contrastive loss between each modality, where the contrastive loss function is InfoNCE. This contrastive loss is directly added to the Focal Loss for backpropagation.
(iii \textbf{GHA}: Employing the proposed GHA Loss, which is similarly combined with the Focal Loss for backpropagation.
In the GHA setting, the number of negative samples $K$ is set to 7, the temperature coefficient $\tau$ is set to 0.005, and the weight $\lambda$ is set to 1. The Dual setting adopts the same values for $K$ and $\tau$ as the GHA setting.
It is important to note that contrastive learning for multi-modal semantic alignment can be implemented in various ways. In this study, we adopt a straightforward approach by directly summing the contrastive and classification losses.
%评价指标
The evaluation metrics used in this study are as follows: top-1 accuracy, denoted as Acc (in percentage \%), the macro kappa value, denoted as $\kappa$, macro precision (P), macro recall (R), and macro F1 score ($F_1$), all expressed in percentage (\%). Additionally, the area under the ROC curve (AUC) is also included as a metric.

\subsection{Real World Dataset}
The performances of the eight baseline models are illustrated in \autoref{c1}, \autoref{radar}, and \autoref{pie}. It is distinctly evident that the Dual and GHA methods, which incorporate the contrastive module, significantly outperform the methods without semantic alignment in every indicator. Meanwhile, as can be observed from \autoref{converge}, the GHA method surpasses the Dual method. The convergence analysis of the model is depicted in Figure 1. Both types of losses steadily decline and approach zero, and simultaneously, the accuracy rate also improves accordingly.

\subsection{Simulated Dataset}
\begin{figure}[htbp] % [h] 固定位置，可替换为 [htb] 等参数
	\centering % 图片居中
	\includegraphics[width=0.5\textwidth]{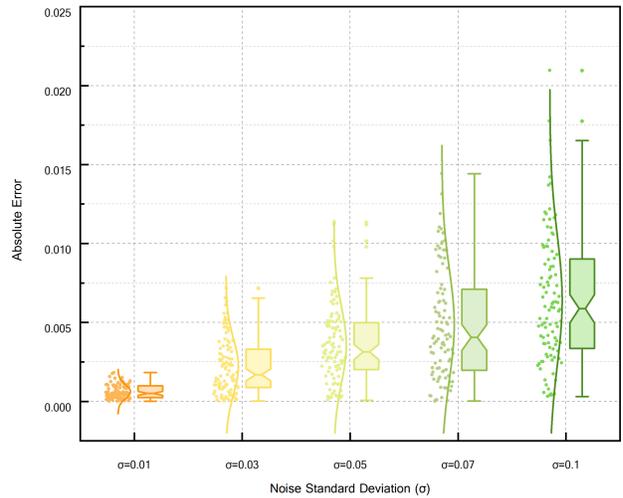}
	\caption{The error distribution of the joint generalized cosine similarity under different noise conditions. The horizontal axis represents the standard deviation of the added white Gaussian noise, and the vertical axis represents the absolute value error between the similarity value of the noisy triplet and the similarity value of the original triplet.} % 图表标题
	\label{noise_robust} % 引用标签
\end{figure}

To evaluate the robustness of the proposed joint generalized cosine similarity (JGCS) to noise interference, we conducted the following experiments:
(i A simulated dataset consisting of 100 samples was generated using a random Gaussian distribution with mean of 0 and standard deviation of 1. Each sample triplet contained three modalities, with each modality having a feature dimension of 256.
(ii The joint generalized cosine similarity for each triplet was computed.
(iii White Gaussian noise with five different standard deviations ($\sigma \in {0.01, 0.03, 0.05, 0.07, 0.1}$) was added to the 100 triplets, and the similarity values for each of the five noisy groups were calculated.
(iv The absolute errors between the similarity values of each noisy group and the original similarity values were computed for all 100 triplets. The distributions of these five groups are presented in \autoref{noise_robust}.
\par As shown in \autoref{noise_robust}, the means of the absolute errors for these five groups of data are as follows: when $\sigma = 0.01$, the mean is 0.0006; when $\sigma = 0.03$, the mean is 0.0022; when $\sigma = 0.05$, the mean is 0.0035; when $\sigma = 0.07$, the mean is 0.0048; and when $\sigma = 0.1$, the mean is 0.0064. These results indicate that the mean absolute error increases monotonically with the standard deviation of the noise, displaying an approximately linear relationship. Given that the joint generalized cosine similarity values fall within the range [0, 1], these errors are considered acceptable. Therefore, we have reason to believe that the proposed similarity measurement method is robust to noise interference.
\par Subsequently, we conducted a simulation experiment for inter-modal contrastive learning across the three modalities. A simulated dataset consisting of 4,000 samples was generated using a random Gaussian distribution with mean of 0 and standard deviation of 1. Each sample triplet contained three modalities, with each modality having a feature dimension of 256. We then designed three encoders with identical structures but independent parameters for each modality. Each encoder is a fully connected neural network with three hidden layers, as described below:
\begin{equation}
	\begin{aligned}
		% 第一层
		\mathbf{h}_1 &= \sigma\left( \mathbf{W}_1^{(\text{mod})} \mathbf{x} + \mathbf{b}_1^{(\text{mod})} \right), \\
		% 第二层
		\mathbf{h}_2 &= \sigma\left( \mathbf{W}_2^{(\text{mod})} \mathbf{h}_1 + \mathbf{b}_2^{(\text{mod})} \right), \\
		% 第三层
		\mathbf{z}^{(\text{mod})} &= \sigma\left( \mathbf{W}_3^{(\text{mod})} \mathbf{h}_2 + \mathbf{b}_3^{(\text{mod})} \right),
	\end{aligned}
\end{equation}
Where $\mathbf{x} \in \mathbb{R}^{256}$ represents the input feature vector of a uni-modality, $\mathbf{h}_1, \mathbf{h}_2 \in \mathbb{R}^{256}$ represents the hidden layer outputs after the first and second linear transformations, respectively. $\mathbf{z}^{(\text{mod})} \in \mathbb{R}^{256}$ represents the final embedding vector for modality.  $\sigma(\cdot)$ represents the ReLU activation function. $\mathbf{W}_1^{(\text{mod})}, \mathbf{W}_2^{(\text{mod})}, \mathbf{W}_3^{(\text{mod})} \in \mathbb{R}^{256 \times 256}$ are learnable weight matrices for modality-specific linear layers and $\mathbf{b}_1^{(\text{mod})}, \mathbf{b}_2^{(\text{mod})}, \mathbf{b}_3^{(\text{mod})} \in \mathbb{R}^{256}$ are learnable bias vectors for modality-specific linear layers.
\par The model is trained using the GHA Loss on these 4,000 samples for a sufficient number of epochs and subsequently tested on the same dataset. The distribution of the feature representations of the modalities in the two-dimensional embedding space is shown in \autoref{distribute}.
\par Part (a) of \autoref{distribute} shows the distribution of seven selected samples after mapping the modalities into the embedding space before training the model. The distributions of the three modalities for the same sample (denoted by the same color) are chaotic. However, after training, for the same seven samples, as shown in part (b) of \ref{distribute}, it is evident that the three modalities of the same color have converged in the embedding space, indicating that semantic alignment has been achieved.
\par Finally, we also compared the computational efficiency of the proposed GHA Loss with the Dual Loss, which calculates loss values pairwise and then performs a summation. The experiments were conducted on a CPU with a frequency of 3.30 GHz.
\par First, we evaluated the computational efficiency of the two loss functions with varying numbers of negative samples. We generated three-modal data with a batch size of 256, where the feature dimension of each modality was 256, and the data followed a Gaussian distribution. The number of negative samples was tested with ten configurations: 5, 10, 15, 20, 25, 30, 35, 40, 45, and 50. Each configuration was run 10 times to calculate the mean and standard deviation, and the results are presented in \autoref{runtime_compare} (a). Next, we assessed the computational efficiency of the two loss functions with varying numbers of modalities. Data with 3 to 12 modalities and a batch size of 256 were dynamically generated. The feature dimension of each modality was 256, and the data followed a standard Gaussian distribution. Each configuration was run 10 times to calculate the mean and standard deviation, with the results shown in \autoref{runtime_compare} (b).
\par It is evident from the figure that, in both cases, the computational efficiency of GHA Loss exceeds that of Dual Loss. In \ref{runtime_compare} (a), when the number of negative samples is 50, the running time of GHA Loss (29.361 ms) is only 63.7\% of that of Dual Loss (46.061 ms). In \ref{runtime_compare} (b), the running time of Dual Loss increases exponentially, while the running time of GHA Loss increases approximately linearly. This disparity occurs because the number of calculations for the pairwise loss function grows according to the combination number as the number of modalities increases.
\section{CONCLUSION}\label{s6}
In this paper, we have identified a significant research gap in the existing literature concerning the calculation of similarities and contrastive losses among multi-modal when the number of modalities exceeds two. To address this issue, we propose a novel joint generalized cosine similarity measurement method, along with the corresponding contrastive loss function, GHA Loss. The proposed method effectively resolves the challenges of chaotic optimization directions and high computational complexity inherent in existing approaches, demonstrating considerable potential. Our experimental results show that GHA Loss offers substantial advantages in terms of superiority, robustness, computational efficiency, and scalability.

\bibliographystyle{IEEEtran} 
%\bibliography{main.bib}

\end{document}